# Metamaterial based broadband engineering of quantum dot spontaneous emission

**Harish N S Krishnamoorthy[1], Zubin Jacob[2], Evgenii Narimanov[2],**

**Ilona Kretzschmar[3] and Vinod M. Menon[1]**

*[1]Laboratory for Nano and Micro Photonics, Department of Physics, Queens College of the City University of New York (CUNY)*
*Tel. (718) 997-3147, Fax: (718) 997-3349, Email: vmenon@qc.cuny.edu*
*[2]Birck Nanotechnology Center, School of Electrical and Computer engineering, Purdue University, West Lafayette, IN 47907, U.S.A*
*[3]Department of Chemical Engineering, City College of the City University of New York (CUNY)*

**Abstract:** We report the broadband (~ 25 nm) enhancement of radiative decay rate of colloidal quantum dots by exploiting the hyperbolic dispersion of a one-dimensional nonmagnetic metamaterial structure.

Control of spontaneous emission is one of the fundamental concepts in the field of quantum optics with applications such as lasers, light emitting diodes, single photon sources among others. The control of emission of quantum dots (QDs) has been reported using photonic crystals and microcavities through the Purcell effect [1-4]. Increasing the photonic density of states (PDOS) is the key to enhancing the spontaneous emission from emitters which have a low quantum yield [5]. There have been several reports on the enhancement of spontaneous emission from QDs embedded in microcavities [3, 4, 6-8]. In all of these demonstrations, the emitter and the emission was confined within the microcavity which enabled the greater interaction between the emitter and the cavity mode. In contrast, the system that we present here does not rely on localization of electromagnetic field for increase in PDOS and thereby the enhancement in spontaneous emission. Instead, the metamaterial structure exploits the hyperbolic dispersion in the PDOS to create more states for the emitter to emit through [9]. One of the key differences in this approach versus the microcavity approach is the broadband nature of the enhancement. In microcavities the enhancement is observed only at the cavity resonant wavelength. In contrast, here the enhancement is observed over the entire emission spectrum of the quantum dots.

The hyperbolic metamaterial is designed using subwavelength layers of metal (silver) and dielectric (titanium dioxide). Effective medium theory [10] for such a layered structure taking into account approximately equal thickness of metal and dielectric layers gives $\varepsilon_{||} = -5 + 0.25i$, and $\varepsilon_{\perp} = 25.1718 + 0.529i$. The opposite signs of the dielectric permittivity tensor in perpendicular directions gives rise to hyperbolic dispersion. The effective medium theory of layered media shows that the hyperbolic dispersion is achieved in a wide spectral range and is tolerant to deviations in the thickness of each of the constituent layers.

The metamaterial consists of alternating layers of silver and titanium dioxide, each of thickness 30nm fabricated via thermal evaporation [Fig. 1(a)]. The top spacer layer is made from either titanium dioxide or silicon nitride. CdSe/ZnS core-shell quantum dots having a peak photoluminescence (PL) emission at 618nm were then deposited on top of the spacer layer. This is done by spin-coating a 25% v/v solution of the QDs followed by curing the sample on a hot-plate. Schematic of the dispersion characteristics of such an anisotropic metamaterial along with that of an isotropic material is shown in Fig. 1(b). For a fixed frequency the wavevector for the anisotropic medium can take on arbitrary large values.

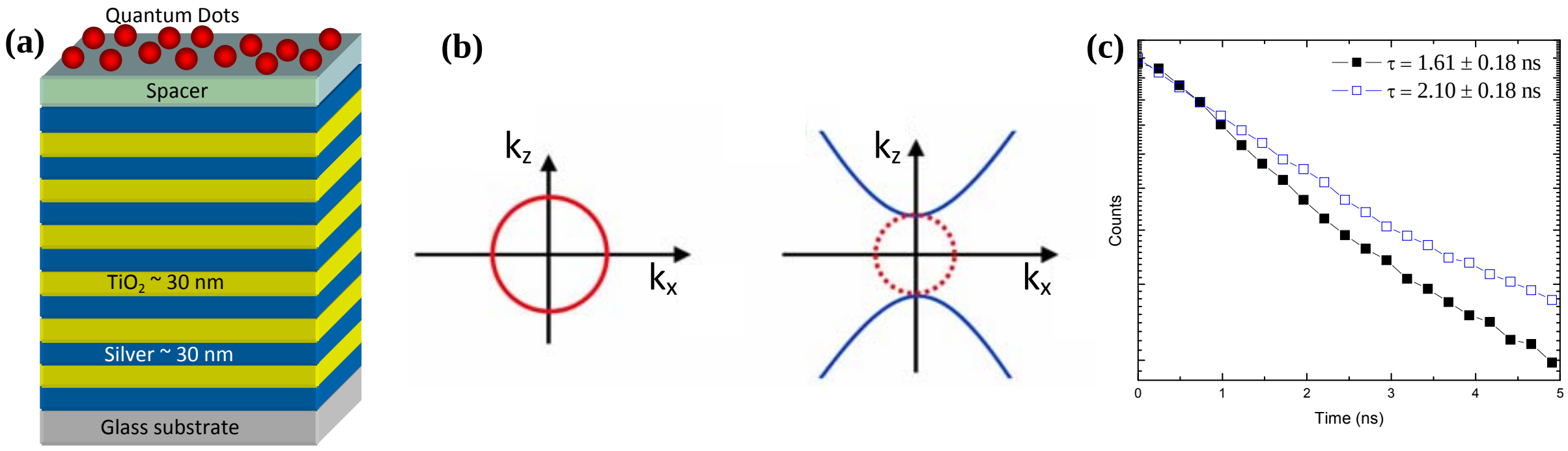

Fig.1 (a) Schematic of the metamaterial structure, (b) Dispersion relation for isotropic medium (left) and for an anisotropic material with hyperbolic dispersion (right). Time resolved PL from the QDs on the metamaterial sample (solid squares) and the reference sample (open squares) with a 35nm thick $TiO_2$ spacer layer.

Time resolved photoluminescence studies were carried out in reflection geometry in order to study the modification in the spontaneous emission of the QDs due to the hyperbolic dispersion in the density of states. The time resolved data for the QDs from the metamaterial sample is compared with the same from a reference sample consisting of a single layer of silver (30nm thick) and the corresponding spacer layer.

Figure 1 (c) shows the time resolved plots for a metamaterial sample with a 35 nm thick titanium dioxide spacer and the corresponding reference. It is observed that the rate of emission in the metamaterial sample is approximately 1.30 times faster than that in the reference sample. These measurements were carried out at the wavelength 618 ± 3nm (peak emission wavelength).

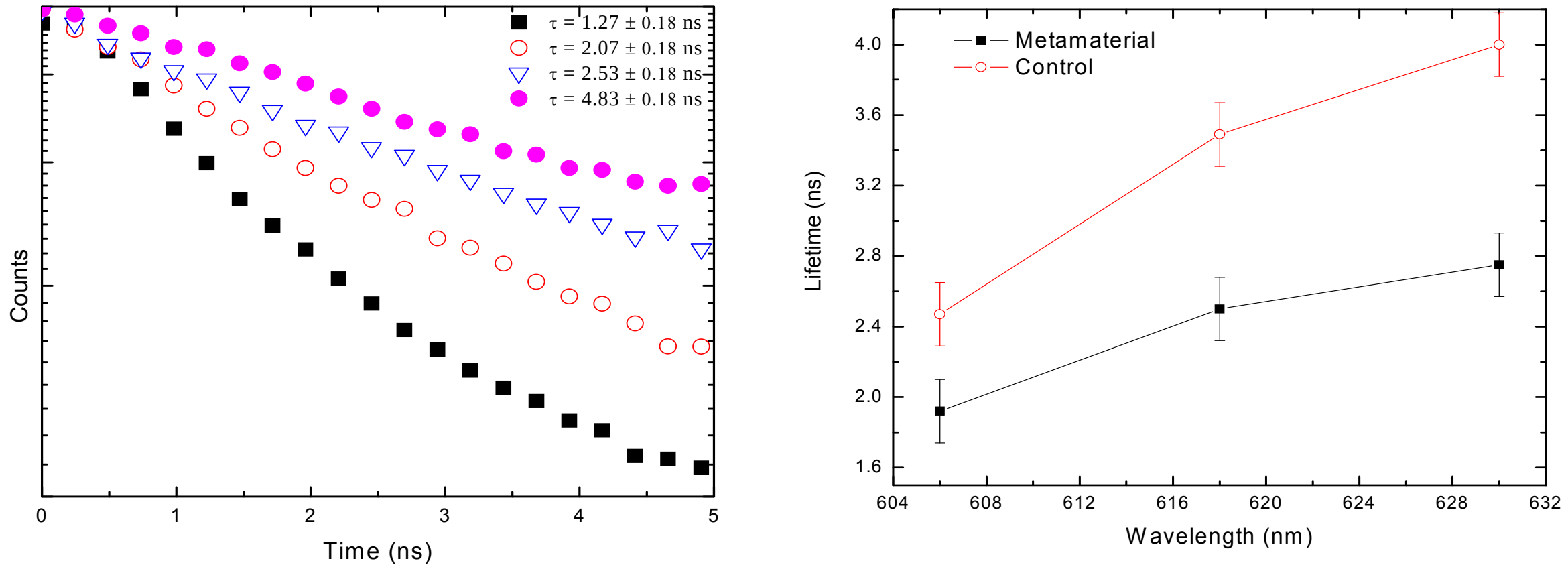

Fig. 2 (a) Time resolved PL as a function of spacer layer thickness: no spacer (solid squares), 35 nm thick $SiN_2$ spacer (open circles), 70 nm thick $SiN_2$ spacer (open triangles), and 95 nm thick $SiN_2$ spacer (solid circles) and (b) Lifetime as a function of wavelength. The measurements were carried out at the emission maximum of 618 nm and at two other wavelengths (± 12 nm) on either side.

Figure 2(a) shows the variation in the lifetime as a function of the spacer layer thickness. The spacer layer was made of silicon nitride using plasma enhanced chemical vapor deposition (PECVD). Clearly, the decay rate is affected by the distance between the emitter and the metamaterial. When the dipole is in close proximity to the metamaterial, most of the emission occurs into the large spatial wave vector channels and this results in a faster rate of emission. In the sample with no spacer (Fig. 2 (a) –solid squares), the quantum dots are sitting on a silver layer and shows much faster decay than the other samples. This decrease in lifetime is partly the contribution of the metamaterial and partly due to the surface plasmon absorption of the silver layer. Simulations for a spacer layer thickness of 30 nm, taking into account the effective medium parameters show a decrease in lifetime of approximately 3 for the metamaterial in comparison to the glass substrate and 1.5 in comparison to the reference sample made of silver. The slight deviation of the theoretical model is related to the intrinsic quantum yield of the dots and deviations from effective medium theory. Finally, we also performed lifetime measurements at two other spectral positions (± 12 nm) on either side of the emission maximum. Results of these measurements are shown in Fig. 2(b) and clearly indicate the broadband nature of the enhancement in the spontaneous emission rate. The increase in lifetime as a function of wavelength is observed for the control sample as well due to the variation in oscillator strength of the inhomogenousely broadened quantum dot ensemble.

We have demonstrated the enhancement in spontaneous emission from quantum dots incorporotaed on a metamaterial sample with hyperbolic dispersion. Unlike in microcavities, we were able to observe this effect over a wide range of wavelengths. The radiative decay rate is found to increase as the dipole is brought closer to the metamaterial.

[1] P. Lodahl, A.F. van Driel, I.S. Nikolaev, A. Irman, K. Overgaag, D. Vanmaekelbergh, W.L. Vos, "Controlling the dynamics of sponatenous emission from quantum dots by photonic crystals", Nature, 430, 654 (2004).

[2] D. Englund, D. Fattal, E. Waks, G. Solomon, B. Zhang, T. Nakaoka, Y. Arakawa, Y. Yamamoto, and J. Vucˇkovic´, " Controlling the spontaenous emission rate of single quantum dots in two-dimensional photonic crystal", Phys. Rev. Lett., 95, 013904 (2005).

[3] M. Bayer, T. L. Reinecke, F. Weidner, A. Larionov, A. McDonald, and A. Forchel, "Inhibition and enhancement of the spontaenous emission of quantum dots in structured microresonators", Phys. Rev. Lett., 86, 3168 (2001).

[4] N. V. Valappil, M. Luberto, V.M. Menon, I. Zeylikovich, T.K. Gayen, J. Franco, B.B. Das, and R.R. Alfano, "Solution processed microcavity structures with embedded quantum dots", Photon. Nanostruct., 5, 184 (2007).

[5] B. Lounis and M. Orrit, "Single-photon sources", Rep. Prog. Phys., 68, 1129 (2005)

[6] G. S. Solomon, M. Pelton, and Y. Yamamoto, "Single mode spontaneous emission from a single quantum dot in a three-dimensional microcavity", Phys. Rev. Lett., 86, 3903 (2001).

[7] J. P. Reithmaier, G. Sek, A. Loffler, C. Hofmann, S. Kuhn, S. Reitzenstein, L. V. Keldysh, V. D. Kulakovskii, T. L. Reinecke and A. Forchel, "Strong coupling in a single quantum dot semiconductor icrocavity system", Nature, 432, 197 (2004)

[8] A. Kiraz, P. Michler, C. Becher, B. Gayral, A. Imamoglu, Lidong Zhang, and E. Hu, "Cavity quantum electrodynamics using a single InAs quantum dot in a microdisk structure", App. Phys. Lett., 78, 3932 (2001)

[9] Z. Jacob, I. Smolyaniniv, E. Narimanov, "Broadband Purcell Effect: Radiative decay engineering with metamaterials", arXiv:0910.3981v2 [physics.optics]

[10] Z. Jacob, L. V. Alekseyev, and E. Narimanov, "Optical hyperlens: Far field imaging beyond the diffraction limit", Opt. Express 14, 8247 (2006).